\documentclass[conference]{IEEEtran}
\IEEEoverridecommandlockouts
\usepackage{cite}
\usepackage{amsmath,amssymb,amsfonts}
\usepackage{algorithmic}
\usepackage{graphicx}
\usepackage{textcomp}
\usepackage{xcolor}
\usepackage{multirow}
\usepackage{booktabs} 
\usepackage{float}
\usepackage{url}
\def\BibTeX{{\rm B\kern-.05em{\sc i\kern-.025em b}\kern-.08em
    T\kern-.1667em\lower.7ex\hbox{E}\kern-.125emX}}
\begin{document}

\title{Quantifying the Relevance of Youth Research Cited in the US Policy Documents\\}

\author{\IEEEauthorblockN{Miftahul Jannat Mokarrama}
\IEEEauthorblockA{\textit{Department of Computer Science} \\
\textit{Northern Illinois University}\\
Dekalb, USA \\
mmokarrama@niu.edu}
\and
\IEEEauthorblockN{Hamed Alhoori}
\IEEEauthorblockA{\textit{Department of Computer Science} \\
\textit{Northern Illinois University}\\
Dekalb, USA \\
alhoori@niu.com}
}

\maketitle

\begin{abstract}
In recent years, there has been a growing concern and emphasis on conducting research beyond academic or scientific research communities, benefiting society at large. A well-known approach to measuring the impact of research on society is enumerating its policy citation(s). Despite the importance of research in informing policy, there is no concrete evidence to suggest the research's relevance in cited policy documents. This is concerning because it may increase the possibility of evidence used in policy being manipulated by individual, social, or political biases that may lead to inappropriate, fragmented, or archaic research evidence in policy. Therefore, it is crucial to identify the degree of relevance between research articles and citing policy documents. In this paper, we examined the scale of contextual relevance of youth-focused research in the referenced US policy documents using natural language processing techniques, state-of-the-art pre-trained Large Language Models (LLMs), and statistical analysis. Our experiments and analysis concluded that youth-related research articles that get US policy citations are mostly relevant to the citing policy documents.
\end{abstract}


\section{Introduction}
In academia, referencing relevant sources is an important aspect of any research. It is a way to give due credit to authors whose research contributes to generating novel research ideas, fosters the ongoing exploration of existing research challenges through advanced tools or techniques, and encourages innovative thinking. It also provides a roadmap of the research process back and forth that is valuable for any researcher to sharpen their background knowledge on the particular research topics \cite{b1, b2}. Moreover, proper citation practices deter plagiarism and enable editors, peer-reviewers, and readers to look deeper into the research topics,  ensure credibility, and add authority to original research \cite{b3}. Given the manifold advantages of citation, the number of citations a research garners has become a widely recognized metric to show the academic excellence of authors and enumerate the impact of their research \cite{b4}. As a result, citations have played a vital role in various decision-making scenarios within academia, such as promotion, research grants, scholarships, and more \cite{b5}.

However, as citations gained recognition and acceptance in academia, the peer pressure to accumulate more citations also accelerated \cite{b6}. While on the one hand, it escalated the research throughputs and competition among peers, on the other hand, the misuse of citations also started to grow more or less. For example, some authors cluttered their reference lists with unnecessary or out-of-context citations, thereby exaggerating the perceived relevance of their findings \cite{b7, b8}. In some cases, these result from unintentional acts of authors, such as when some editors or reviewers request authors to cite a certain percentage of previously published work from their journal to get the research accepted \cite{b6}. In some other cases, it might be intentional, such as adding the names of authors who did not contribute to the work or citing authors whose work is irrelevant to the citing research \cite{b9,b10}. Such acts hinder the proper recognition of true researchers and research and raise concerns over academic integrity and ethical citation practices \cite{b3}.

Despite these controversies, the citation count continues to be regarded as an inevitable metric for measuring the academic impact of the research \cite{b11}. However, this metric limits the scope of research impact and has several limitations, including the need for time to accumulate and differences between disciplines. Therefore, researchers have been exploring the use of supplementary metrics (e.g., altmetrics \cite{b12,b13,b14}), not necessarily as alternatives but rather as additional parameters that can provide insights into the impact of research on society \cite{b15,b16, b17, b18}. Recently, referencing research in policy documents has gained more attention, highlighting the societal impact of research. 

The Centers for  Disease Control and Prevention (CDC) describes the policy as “a law, regulation, procedure, administrative action, incentive, or voluntary practice of governments and other institutions” \cite{b19}. In contrast, according to the National Academies of Science, the purpose of the research is “to extend human knowledge of the physical, biological, or social world beyond what is already known.” \cite{b20}. Although research plays a crucial role in shaping policy, there is no clear evidence that all research cited in policy documents is relevant. This raises concern because citations in policy documents have been acknowledged as a potential social impact measurement metric for research \cite{b21}. Therefore, invalidated research citations in policy documents can undermine the trustworthiness of policymakers and nullify their public acceptance. Moreover, \textbf{it can lead to unintended policy outcomes, particularly when addressing marginalized or underrepresented groups \cite{b22}, such as youth, who often lack a voice in policymaking \cite{b23} though constituting one-fifth of the US population \cite{b24}}. 

It is crucial that policies intended for youth are informed by research that accurately reflects the needs and experiences of young people. However, when irrelevant or selectively chosen citations are used, policies may emphasize outdated or misaligned perspectives, failing to address the specific issues youth face. This can result in policies that do not adequately support youth development or participation, limiting opportunities for meaningful engagement and inclusion in policy decisions. Furthermore, biased citations can hinder innovation in youth policy by reinforcing traditional approaches and stifling evidence-based advancements that address the evolving challenges younger generations face. Therefore, \textbf{it is essential to ensure that relevant research is cited in youth policy documents to foster more responsive, equitable, and future-oriented policies}.

Currently, there is limited research that examines the relevance between cited research articles and citing policy documents following a quantitative approach. However, the unprecedented advancement in computational science could offer a sophisticated framework for such analysis. For example, the high-level computational power of Large Language Models (LLMs) pretrained with a massive collection of datasets with complex deep learning architecture has shown remarkable performance within several Natural Language Processing (NLP) tasks such as text generation, text summarization and sentiment analysis. The impressive human-like text processing capabilities of such pre-trained transformer-based models have enabled researchers to deploy them in building their models without even fine-tuning \cite{b25}. Therefore, in this research, we aim to use a quantitative approach to unfold the following research questions: 

\begin{itemize}
    \item \textbf{RQ1:} \textit{Can we use \textbf{pretrained LLMs} for analyzing research and policy Documents?}
    \item \textbf{RQ2:} \textit{Do policymakers use \textbf{relevant} research evidence in policy documents? }
\end{itemize}

We focused on youth research that is cited in the US policy documents for this experimental analysis. In section II we provided the related work and in sections III and onwards, we provided a detailed description of our whole research and analysis process.

\section{Related Work}
\subsection{Citation Misuse in Academia}
Several studies discussed the prevalence of citation errors and the reasons for such errors in academia. Greenberg \cite{b26} used 242 PubMed-indexed papers with 675 citations to construct a biomedical citation network to analyze the possible citation errors around a publication belief system. They specifically defined several forms of citation errors, including citation bias, distortion, amplification, and invention, that could be used to identify the nature of citation errors in research. In \cite{b27}, Agarwal et al. discussed several forms of citation errors, namely, non-citing error, factual error, selective citation, incorrect source type, insufficient support, wrong citation, and incorrect technical details. They stated that citation errors emerge due to author-related, article-related, journal-related, and guidelines-related factors, and it is imperative to go through the full text of research to avoid citation errors.

Wilhite and Fong \cite{b28} investigated another form of citation misuse-coercive citation, which implies directly or indirectly pressuring authors to add citations from the editor’s journal to increase the impact factor. In another study \cite{b9}, they surveyed 110K scholars from multiple disciplines and discovered from the responses that citation misattribution in research proposals and publications is prevalent in academia, with notable differences by co-authors, academic level, genders, research area, and publication record. Stelmakh et al. \cite{b29} examined around 1.5K papers and 2K reviewers of two conferences on machine learning and algorithmic economics to verify whether there is any citation bias prevalent in the peer review process. Unsurprisingly, they found that citing the reviewer's work in the submitted papers increased the acceptance score by a value of 0.23 on a 5-point Likert item, providing notable evidence of citation bias. 

Meyer \cite{b30} examined the significance of reference citations, typical issues, and solutions implemented by publishers. He inferred that references offer a way to quantify the impact of research or research repositories, and therefore, any issues with reference authenticity have the potential to downplay the significance of a journal article or an author's research. Pavlovic et al. \cite{b31} investigated inaccurate citations in citing papers using around 5K articles in the field of biomedicine. They considered a citation accurate if the referenced article is valid and a similar assertion was made by the citing writers as stated in the source. They found at least one inaccurate citation in 11 papers, and 38.4\% of the papers cited fictional results, while 15.4\% cited fabricated results claimed to be from the referenced articles. Dumas-Mallet and Gonon \cite{b32} investigated how public health is affected by citation misuse and found that in the field of biomedicine, over 10\% of citations are absolutely wrong and deceive the readers.
\subsection{Use of Citation in Policymaking}
Researchers have increasingly explored the extent of research citation in policy documents in recent years as the urge to make evidence-based policy has been growing. However, very few studies have examined the citation patterns of research in policy documents \cite{b33}. In \cite{b34}, Mirvis discussed the significance of such investigation and proposed that creating efficient methods for combining research results with the context that policymakers need to know could be a potential field of exploration. Vilkins and Grant \cite{b35} argued that incorporating scientific papers into policy documents is contingent upon the institutional attitudes and practices surrounding their utilization. They classified the utilization of scientific knowledge in policy-making into three distinct purposes: 1) ‘Instrumental,’ where the contribution is straightaway and quantifiable; 2) ‘Conceptual,’ where the impact is long-term and consequential; and 3) ‘Symbolic,’ where research usage is selective and favors the political standpoint. 

Malekinejad et al. \cite{b36} mentioned the possibility of evidence used in policy being manipulated by individual, social, or political biases that may lead to the presence of inappropriate, fragmented, or archaic research evidence in policy. Zardo and Collie \cite{b37} analyzed the references used in transport injury compensation policy documents in Victoria, Australia, and found that all the references were used in support of policies with minimal utilization of research evidence. A content analysis of the Australian childhood obesity prevention policy documents conducted by Newson et al. \cite{b21} found that approximately half (47\%) of the references in those documents were related to obesity research. They also noted that most of this research was peer-reviewed (around 75\%) and served to validate claims within the policy documents. Yu et al. \cite{b38} analyzed 885 policy documents and scholarly mentions within those documents and found that the most common way scholarly evidence is mentioned is through formal references. They also found that nearly one-third of scholarly mentions are used to summarize research findings, and around one-third are used to support policy claims.

From the above background research, we can say that it is necessary to find a quantitative way to check whether policymakers use \textbf{relevant} research evidence in policy documents and, if yes, to what extent. With this aim, in the following sections, we will describe how we can find the relevance between research and citing policy documents using state-of-the-art pretrained open-source LLMs available in Hugginface\footnote{https://huggingface.co/}.

\section{Dataset preparation}
\subsection{Dataset Collection}
Our data collection process started by collecting research articles cited by the US policy documents between January 1, 2000, and December 31, 2022, from the online policy document repository \textit{Overton}\footnote{https://www.overton.io/}. \textit{Overton} is the largest policy data source that records the citation data of policy documents worldwide through a wide range of collaboration with stakeholders. 
Since it was not possible to categorize and search research articles based on age (which is a number) or range of ages to indicate research on youth, we aimed our search for research articles on three primary topics: `Child,' `Teen,' `Youth,' and other relevant topics. We picked thirty keywords from the website RelatedWords\footnote{https://relatedwords.org} that are relevant to those topics to filter data from \textit{Overton}, as shown in Table \ref{tab:keywords}.

\begin{table}[htbp]
  \caption{Keywords Chosen from RelatedWords for Filtering Research Articles from \textit{Overton} that Align with “Child, Teen, and Youth”}
  \label{tab:keywords}
  \centering
  \begin{tabular}{@{}lp{7cm}@{}} 
    \toprule
    \textbf{Topics} & \textbf{Keywords} \\
    \midrule
    Child & baby(-ies), kid, child(-ren), caregiver AND child, childhood, newborn, infant, toddler \\
    Teen & adolescent, adolescence, boy, girl, juvenile, teenage, teen \\
    Youth & adulthood, caregiver AND youth, youth, young \\
    Others & bully, college, foster, kindergarten, parent, preschool, school, stepchild, student \\
    \bottomrule
  \end{tabular}
\end{table}

Given the primary criteria to search in \textit{Overton} (i.e., timeline from 2000 to 2022, citing policy region: `US', and document type: `research articles'), the data filtering was done as follows:

\begin{itemize}
    \item For each keyword, data was collected from \textit{Overton} using three additional search criteria: \textit{published date}, \textit{relevance}, and \textit{citations}, each criterion at a time.  
    \item Since a search on \textit{Overton} does not return data for more than 250 research articles, our data collection criterion was further split on a yearly basis.
\end{itemize}
Thus, we had the coverage of almost all research articles in the \textit{Overton} dataset for each keyword. 
Next, we merged all the data files collected separately into a single file and used them upon further preprocessing stages as described in the next subsections.

\subsection{Dataset Filtering}
\begin{figure}[htbp]
    \centerline{\includegraphics[width=1\linewidth]{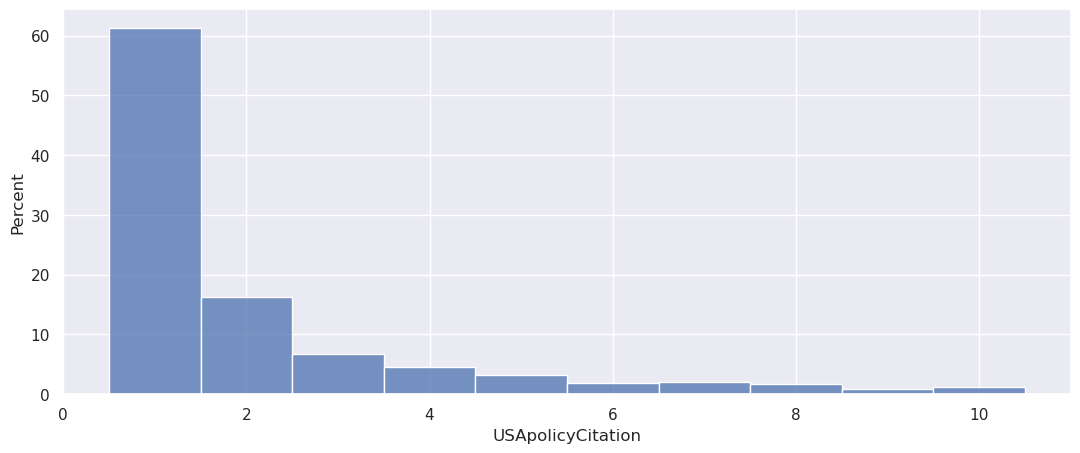}}
    \caption{Distribution of the citation count of research articles (1 to 10) in the US policy documents.}
    \label{fig:density}
\end{figure}

Given the likelihood of encountering duplicate research articles, we filtered the dataset by removing duplicate research articles, missing and inconsistent publication and policy citation dates, and missing research article or policy document titles. Next, we looked into the distribution of the dataset based on US-based policy citation count. We found that less than 10\% of the research articles received citations beyond 1 to 10 in the US during the selected years. Therefore, we filtered those research articles and selected the rest 52,279 research articles that received citations ranging from 1 to 10 within 37,620 US policy documents (Figure \ref{fig:density}).

\subsection{PDF Collection}
At this stage, we collected PDFs for both research articles and citing policy documents. Here is the summary of our PDF collection process for both document types:
\begin{itemize}
\item{\verb|Research articles|}: We extracted the research articles' PDFs using Digital Object Identifiers (DOIs). In total, we had PDFs for 5,884 research articles out of 52,279 (11.25\%) that have citations in 4,394 policy documents.
\item{\verb|Policy documents|}: \textit{Overton} comes with direct PDF links of the citing policy documents, and therefore, for each of the 5,884 research articles, we extracted PDFs of citing policy documents by web-scraping using the Python \textit{AsyncHTMLSession}\footnote{https://requests.readthedocs.io/projects/requests-html/en/latest/} library. This library is more useful in such tasks than the other widely used packages like \textit{SELENIUM} \footnote{https://selenium-python.readthedocs.io/} and \textit{Beautiful Soup}\footnote{https://pypi.org/project/beautifulsoup4/} as its support of scraping from JavaScript web pages enables more coverage of the web documents. In this process, if the PDF of a research article was faulty, or none of the PDFs of the citing policy documents could be extracted, or all of the citing policy document PDFs were found defective, we removed that research article record. Thus we have 2,301 research article records and corresponding 2,818 citing policy documents.
\end{itemize}

\subsection{Text Extraction}
We considered three cases for conducting our experiments, as given below: 
\begin{itemize}
    \item{\verb|Case 1|}: \textbf{Titles} of the research articles and \textbf{titles} of the policy documents. 
    \item{\verb|Case 2|}: \textbf{Abstracts} of the research articles and \textbf{full-text} of the policy documents.
    \item{\verb|Case 3|}: \textbf{Full-text} research articles and \textbf{full-text } of the policy documents.
\end{itemize}

The reason for using `full-text' instead of `abstract' of the policy documents in {\verb|Case 2|} is that policy documents are diverse in their structures\footnote{https://politicalscienceguide.com/home/policy-paper/} and may even follow inconsistent formats within a policy source. Therefore, instead of looking for an abstract, we used full-text of the policy documents. 

Since we already have the titles of articles and citing policy documents as an attribute in our preprocessed \textit{Overton} dataset, we used the title texts of the 2,301 research articles and the corresponding 2,818 policy documents for case 1. 

To extract the abstracts of the research articles, we used the Python API \textit{ScienceParse}\footnote{https://pypi.org/project/science-parse-api} and converted each not-null abstract text into a list of sentences using the \textit{Natural Language Toolkit} (\textit{NLTK})\footnote{https://www.nltk.org/} sentence tokenizer. 

For collecting the full texts of research articles and policy documents, we converted the PDFs into texts using the Python library \textit{PyMuPdf}\footnote{https://pypi.org/project/PyMuPDF} and did primary preprocessing (e.g., removing emails, URLs, numbers, and so on) on those texts. 

Finally, for all cases, we tokenized the texts into a list of sentences and removed punctuation, leading, and trailing whitespaces in each sentence.

\section{Relevance Score Calculation}
We used TF-IDF vectorizer to generate the document-term matrix and 11 different transformer-based pretrained Large Language Models (LLMs) within the Sentence-BERT (SBERT) \cite{b39} framework to generate sentence embeddings of the preprocessed texts. Sentence embeddings play a pivotal role in enhancing the capabilities of information retrieval systems, particularly by improving semantic search. This technique is a core component of natural language processing (NLP) that involves transforming entire sentences into fixed-size numerical vectors that encapsulate their semantic meaning. These embeddings facilitate the comparison of sentences based on their semantic similarity, enabling the identification of relationships between sentences beyond mere keyword matching. By converting both queries and documents into vector representations, it can assess semantic similarity using metrics such as cosine similarity, which allows for the retrieval of semantically relevant documents rather than those that only match specific terms. In this work, we used SBERT among various sentence embedding models since it is widely recognized for its superior performance in semantic textual similarity tasks.

Table \ref{tab:models1} briefly overviews the pretrained LLMs used for generating embedding of the texts in our experiment. The selection of the models was according to the following criteria:
\begin{enumerate}
    \item We included the bert-large-uncased model (model 2 in Table \ref{tab:models1}) as the base model.
    \item Next, we searched for other transformer models in the HugginFace using the following criteria: ‘sentence similarity’ as the task, ‘English’ as the language, and sorting based on ‘most likes’ and ‘most downloads’. We had models 3-6 in Table \ref{tab:models1} according to these criteria. 
    \item In addition to these, we have considered some pretrained language models that are trained on a vast amount of documents relevant to the field of interest. Models 7-12 in Table \ref{tab:models1} follow that criterion.
\end{enumerate} 

\begin{table*}[htbp]
    \caption{Bert models used in SBERT framework for sentence embeddings.}
    \label{tab:models1}
    \centering
    \begin{tabular}{clll}
        \toprule
        \textbf{No.} & \textbf{Models} & \textbf{Pretraining} & \textbf{Fine-Tuning}\\
        \midrule
        M2 & bert-large-uncased & BookCorpus+English Wikipedia & N/A \\
        & & (excluding lists, tables, and headers)  & \\
        M3 & all-distilroberta-v1 & distilroberta-base model  & 1B sentence pairs from \\
        & & & multiple datasets \\
        M4 & all-mpnet-base-v2 & mpnet-base model & 1B sentence pairs from \\
        & & & multiple datasets \\
        M5 & all-MiniLM-L6-v2 & MiniLM-L6-H384-uncased model  & 1B sentence pairs from \\
        & & & multiple datasets \\
        M6 & sentence-t5-xl & 2B Q/A from Web Forums & Stanford Natural Language \\
        & & & Inference (SNLI) dataset \\
        M7 & bert-political-election & 5M English tweets about the 2020 & N/A \\
        & 2020-twitter-mlm & US Presidential election   & \\
        M8 & legal-bert-base-uncased & Legal documents like legislation, & N/A \\
        & & court cases, contracts, and so on & \\
        & & from the UK and USA  & \\
        M9 & specter & More than 6M triplets of scientific & N/A \\
        & & paper citations  & \\
        M10 & scibert\_scivocab\_uncased & 1.14 M full text scientific papers from & N/A \\
        & & Semantic Scholar  & \\
        M11 & Bio\_ClinicalBERT & All MIMIC notes (electronic health records) & N/A \\
        M12 & biobert-base-cased-v1.2  & English Wikipedia (Wiki), BooksCorpus & N/A \\
        & & (Books), PubMed abstracts (PubMed), & \\
        & & and PMC articles (PMC) & \\
        \bottomrule
  \end{tabular}
\end{table*}

\begin{figure}
    \centering
    \includegraphics[width=0.85\linewidth]{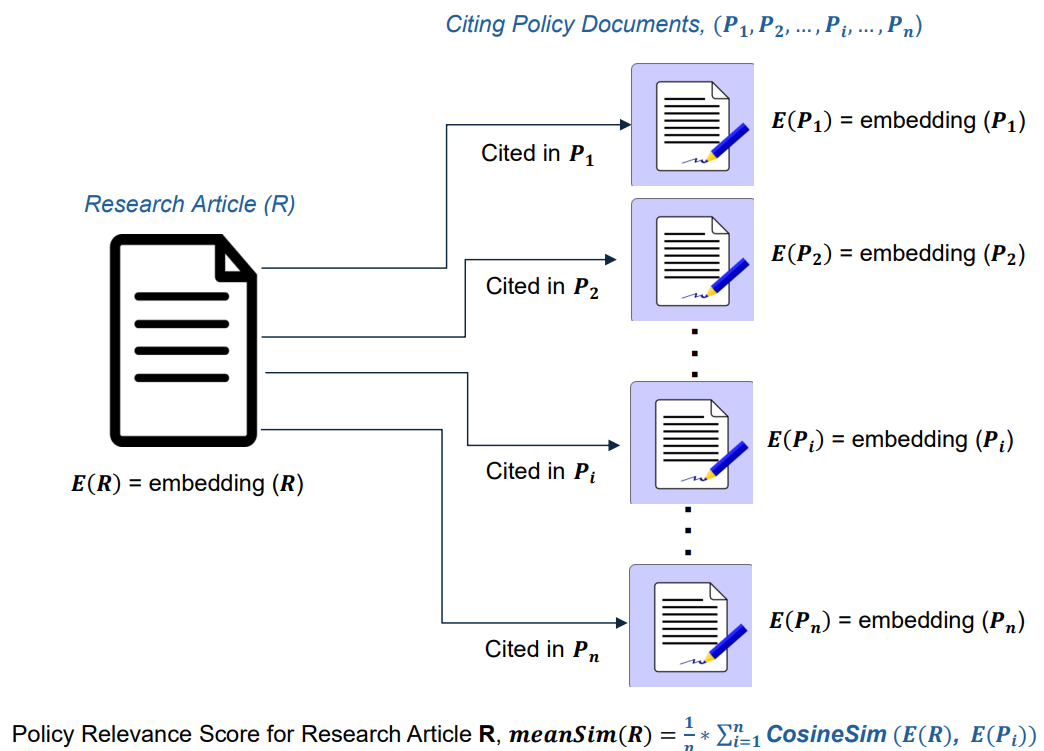}
    \caption{Research article and citing policy document relevance score calculation }
    \label{fig:relevance}
\end{figure}

To get the policy relevance score for each research article, we calculated the \textit{\textbf{cosine}} similarities of embeddings (TF-IDF vectorized output for TF-IDF vectorizer) for each research article and corresponding citing policy documents. Finally, we averaged that similarity values of each research article with the corresponding citing policy document(s) and returned that score as our final relevance score for that particular research article. In Fig \ref{fig:relevance}, the relevance score calculation process is shown. We followed the process across all models in every case for each research article. Thus, we had three updated datasets corresponding to three cases with 12 new attributes in each dataset for 11 LLMs and TF-IDF vectorizer holding the policy relevance scores for each research article.

Finally, we removed the records from each of those three datasets for which there were \textit{null} relevance scores and got the updated datasets with records for \textbf{1,890}, \textbf{2,008}, and \textbf{2,280} research articles for {\verb|Case 1|}, {\verb|Case 2|}, and {\verb|Case 3|}, respectively. 

\section{Model Analysis and Discussion}
We utilized descriptive and inferential statistics to experiment and analyze the distribution of relevance scores returned using a TF-IDF vectorizer and different models in each dataset.

\begin{figure*}[htbp]
\centering
\includegraphics[width=0.85\linewidth]{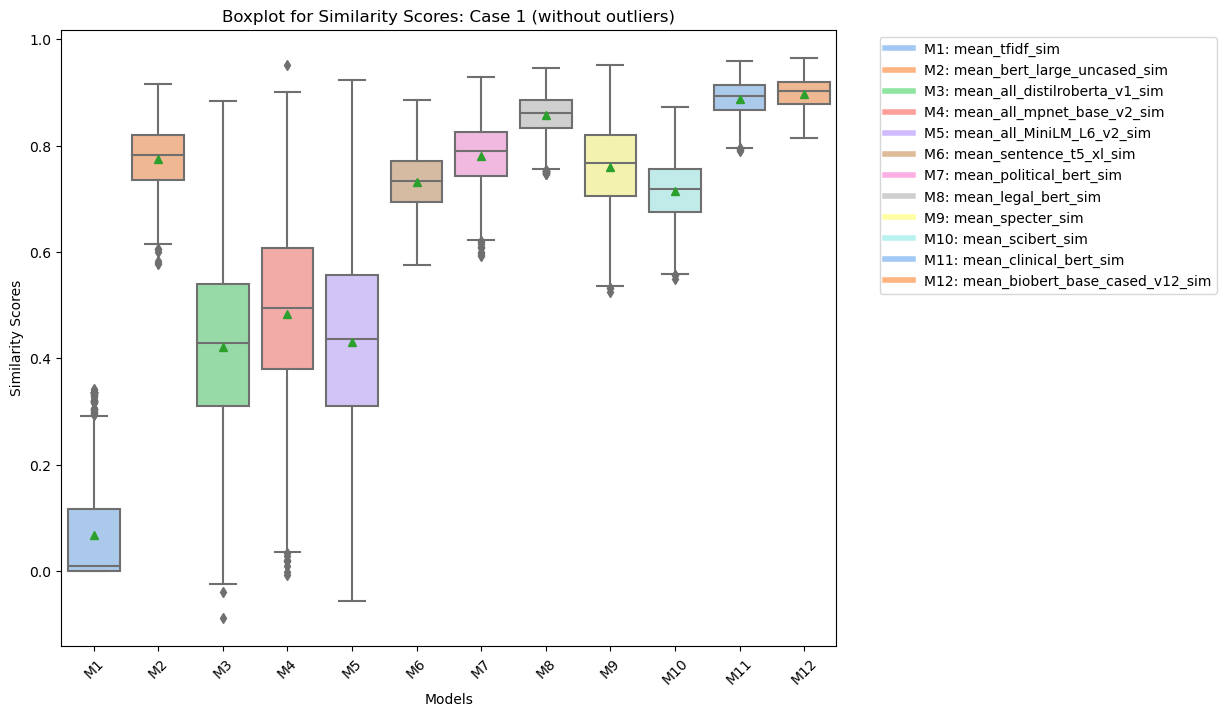}
\caption{Box-Whisker plots of the relevance scores for all models in \textbf{Case 1} after removing outliers}
\label{fig:case1}
\end{figure*}

\begin{figure*}[htbp]
\centering
    \includegraphics[width=0.85\linewidth]{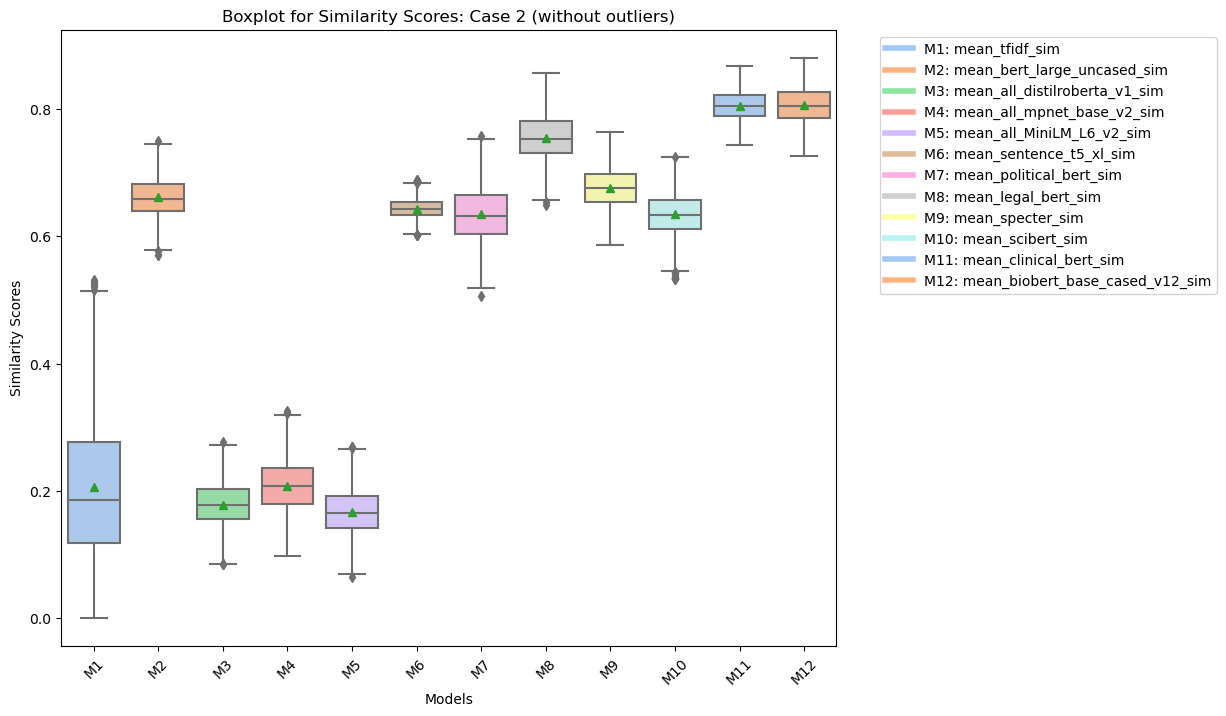}
    \caption{Box-Whisker plots of the relevance scores for all models and in \textbf{Case 2} after removing outliers}
    \label{fig:case2}
\end{figure*}

\begin{figure*}[htbp]
\centering
    \includegraphics[width=0.85\linewidth]{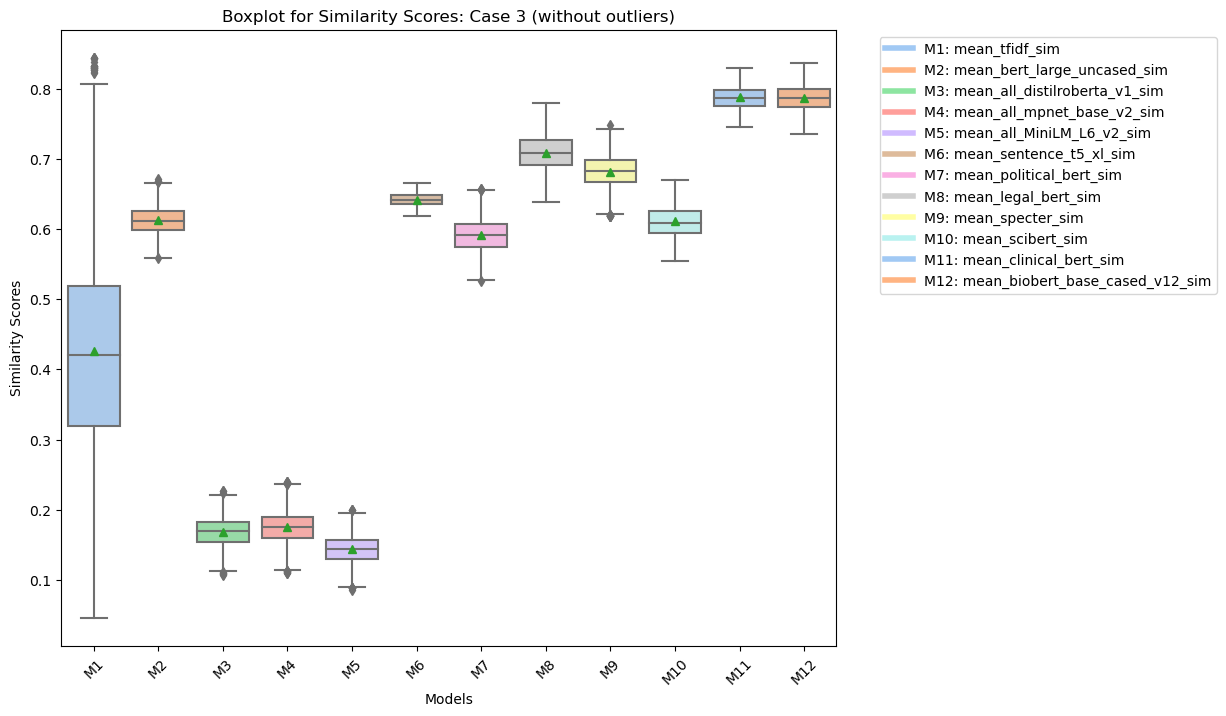}
    \caption{Box-Whisker plots of the relevance scores for all models and in \textbf{Case 3} after removing outliers}
    \label{fig:case3}
\end{figure*}

To examine the descriptive statistical nature of the relevance scores of our datasets, we used the box and whisker plot that can show the distribution of data, maximum value, minimum value, measures of central tendency- mean and median, measures of variability- range and interquartile range (IQR), and outliers (relevance scores that are beyond 1.5 times IQR) if there is any. As we noticed some outliers in each case, we removed those records of research articles from the datasets and got the data distribution as shown in Figure \ref{fig:case1}, Figure \ref{fig:case2}, and Figure \ref{fig:case3} for {\verb|Case 1|}, {\verb|Case 2|}, and {\verb|Case 3|}, respectively. In this way, we got \textbf{1,607} (85\%), \textbf{1,638} (82\%), and \textbf{1,997} (88\%) records of research articles for each case, respectively.


As we inspected the plots of Figure \ref{fig:case1}, Figure \ref{fig:case2}, and Figure \ref{fig:case3}, we had some noteworthy observations. They are discussed below.
\begin{enumerate}
   \item For all three cases, we observed four clusters of models according to the closeness of the distribution of the relevance scores of research articles. Those are:
\begin{itemize}
        \item {\textbf{\textit{cluster1}}: mean\_tfidf\_sim (M1)} 
        \item {\textbf{\textit{Cluster2}}: mean\_all\_distilroberta\_v1\_sim (M3), mean\_all\_mpnet\_sim (M4), mean\_MiniLM\_sim (M5)}
        \item {\textbf{\textit{Cluster3}}: mean\_bert\_large\_uncased\_sim(M2), mean\_sentence\_t5\_xl\_sim(M6), mean\_political\_bert\_sim(M7), mean\_legal\_bert\_sim(M8), mean\_specter\_sim(M9), and mean\_scibert\_sim(M10)}
       \item {\textbf{\textit{Cluster4}}: mean\_clinical\_bert\_sim (M11), and mean\_biobert\_sim (M12)}
    \end{itemize}

    \item TF-IDF technique had an increased value of relevance scores as the text length increased from {\verb|Case 1|} to {\verb|Case 3|}. For transformer-based models, which consider the semantics of the texts, the scores decreased to different degrees.
    \item Models in \textbf{\textit{cluster4}} show the highest relevance scores in all cases and lower variability of relevance scores across different cases than other models.
    \item For pretrained LLMs, in {\verb|Case 3|}, the decrease of relevance scores from {\verb|Case 2|} was minimal compared to the decrease from {\verb|Case 1|} since in {\verb|Case 1|}, we only consider short text (titles).
   \item The distribution of the relevance scores for models in {\verb|Case 2|} went through a notable shrinkage in {\verb|Case 2|} and {\verb|Case 3|} than in {\verb|Case 1|} when compared to other models.
\end{enumerate}

     We also calculated \textit{Spearman's rank correlation}\footnote{https://docs.scipy.org/doc/scipy/reference/generated/scipy.stats.spearmanr.html} to find the instance-based highly similar models' groups. It is a non-parametric measure of the strength and direction of association between two variables that looks at whether an increase in one variable tends to be associated with an increase or decrease in another variable without requiring the relationship to be linear. Using this technique, we listed the group of models that showed a correlation greater than or equal to +0.70 and less than or equal to -0.70. The groups of similar models found in each case are listed in Table \ref{tab:model cluster}. It is worth noting that we did not find any model that showed negative correlation with another model.

\begin{table*}[htbp]
  \caption{Groups of similar models in each case}
  \centering
  \begin{tabular} {p{1cm}p{4.9cm}p{4.9cm}p{4.9cm}}
  \toprule
 \textbf{Groups} & \textbf{Case 1} & \textbf{Case 2} & \textbf{Case 3}  \\ \midrule  
  {Group 1} & mean\_tfidf\_sim (M1) & mean\_tfidf\_sim (M1) & mean\_tfidf\_sim (M1) \\
  \midrule
   \multirow{4}{*}{Group 2} & mean\_all\_distilroberta\_v1\_sim (M3) & mean\_all\_distilroberta\_v1\_sim (M3) & mean\_all\_distilroberta\_v1\_sim (M3)\\ 
   &mean\_all\_mpnet\_base\_v2\_sim (M4) &mean\_all\_mpnet\_base\_v2\_sim (M4) &mean\_all\_mpnet\_base\_v2\_sim (M4) \\
   & mean\_all\_MiniLM\_L6\_v2\_sim (M5) & mean\_all\_MiniLM\_L6\_v2\_sim (M5) & mean\_all\_MiniLM\_L6\_v2\_sim (M5)\\
   & mean\_sentence\_t5\_xl\_sim (M6)\\

 \midrule

  {Group 3} & mean\_specter\_sim (M9) & mean\_specter\_sim (M9) & 
  mean\_specter\_sim (M9)
   \\
  \midrule
  \multirow{3}{*}{Group 4} & \textbf{mean\_bert\_large\_uncased\_sim (M2)}& \textbf{mean\_bert\_large\_uncased\_sim (M2)} & 
  \textbf{mean\_bert\_large\_uncased\_sim (M2)}  \\
  & mean\_political\_bert\_sim (M7) & mean\_political\_bert\_sim (M7) & \textbf{mean\_political\_bert\_sim (M7)} \\
   & mean\_legal\_bert\_sim (M8) & mean\_legal\_bert\_sim (M8) & \textbf{mean\_legal\_bert\_sim (M8)}\\
   & & mean\_clinical\_bert\_sim (M11) & mean\_clinical\_bert\_sim (M11)
  \\
    &  & mean\_biobert\_base\_cased\_v12\_sim (M12) & mean\_biobert\_base\_cased\_v12\_sim (M12)  
   \\
   
  \midrule
  {Group 5} & mean\_scibert\_sim (M10) & mean\_sentence\_t5\_xl\_sim (M6)  & mean\_sentence\_t5\_xl\_sim (M6)\\
  \midrule
  \multirow{5}{*}{Group 6} & mean\_political\_bert\_sim (M7)  & \textbf{mean\_scibert\_sim (M10)}   & \textbf{mean\_scibert\_sim (M10)} \\
  &  mean\_legal\_bert\_sim (M8) & mean\_political\_bert\_sim (M7) & mean\_clinical\_bert\_sim (M11)\\
  & \textbf{mean\_clinical\_bert\_sim (M11)}  &  mean\_legal\_bert\_sim (M8) & mean\_biobert\_base\_cased\_v12\_sim (M12)\\
  & \textbf{mean\_biobert\_base\_cased\_v12\_sim} & mean\_clinical\_bert\_sim (M11) &\\
  &\textbf{(M12)} &mean\_biobert\_base\_cased\_v12\_sim (M12) & \\
  \bottomrule
\end{tabular}
\label{tab:model cluster}
\end{table*}

From the table, we observe almost similar grouping patterns of the models as observed from Figure \ref{fig:case1}, Figure \ref{fig:case2}, and Figure \ref{fig:case3}. In {\verb|Case 1|}, although the models mean\_clinical\_bert\_sim (M11) and mean\_biobert\_base\_cased\_v12\_sim (M12) demonstrate higher relevance both with each other and with the other models mean\_political\_bert\_sim (M7) and mean\_legal\_bert\_sim (M8) (Group 6), they show lower relevance (less than 0.70) with model mean\_bert\_large\_uncased\_sim (M2) (Group 4). Consequently, they are categorized into separate groups. Similarly, in {\verb|Case 2|}, though model mean\_scibert\_sim (M10) shows higher relevance with other models mean\_political\_bert\_sim (M7), mean\_legal\_bert\_sim (M8), mean\_clinical\_bert\_sim (M11), and mean\_biobert\_base\_cased\_v12\_sim (M12) (Group 6), it shows lower relevance with model mean\_bert\_large\_uncased\_sim (M2) (Group 4), and therefore, placed in different groups (Group 4 and Group 6). Finally, in {\verb|Case 3|}, model mean\_scibert\_sim (M10) shows higher relevance with two other models mean\_clinical\_bert\_sim (M11), and mean\_biobert\_base\_cased\_v12\_sim (M12) (Group 6). However, it shows lower relevance than the threshold with models mean\_bert\_large\_uncased\_sim (M2), mean\_political\_bert\_sim (M7), and mean\_legal\_bert\_sim (M8) (Group 4). Therefore, they are grouped separately in Group 4 and Group 6.

If we examine the observations above from the pretraining and finetuning datasets of the models (Table \ref{tab:models1}), we can somewhat assume the reasons for the model's behavior. For example, all the models in \textbf{\textit{cluster2}} were fine-tuned with the same dataset, and the distribution of the relevance scores for those models was the lowest compared to all other clustered models and even TF-IDF relevance scores, which don’t consider the semantics of the text. 

In the case of \textbf{\textit{cluster3}}, the datasets were pretrained on a large volume of text documents more likely to be related to public interests. Again, from Table \ref{tab:models1}, we found that the type of datasets used to pre-train the models in \textbf{\textit{cluster4}} were primarily related to healthcare, and there is evidence from previous research that research related to healthcare gets more citations in policy documents than other fields \cite{b40, b41}. Our experiment with youth research cited in policy documents also follows a similar alignment. 

To be more confident about the conclusions stated in the previous section and generalize for the population, we conducted a one-sample and upper-tail \textit{t}-test on our sample dataset of relevance scores with a 5\% significance level. The resulting statistically significant models are listed in Table \ref{tab:inferential statistical analysis}. We observed that \textbf{M8} (mean\_legal\_bert\_sim), \textbf{M11} (mean\_clinical\_bert\_sim), and \textbf{M12} (mean\_biobert\_sim) are statistically significant in all three cases.

\begin{table*}[htbp]
  \caption{Different cases and models showing statistically significant results.}
  \centering
  \begin{tabular} {p{1cm}p{5cm}p{1cm}p{1.2cm}p{1.8cm}p{1.6cm}p{3.2cm}}
  \toprule
 \textbf{Cases} & \textbf{Models} & \textbf{Mean} & \textbf{Median} & \textbf{p-values} & \textbf{t-scores} & \textbf{95\% confidence interval} \\ \midrule  
  \multirow{8}{*}{Case 1} & \textbf{mean\_biobert\_base\_cased\_v12\_sim (M12)} & 0.897& 0.786 & 0.0 & 260.581 & [ 0.895 , 0.898 ]  \\ 
 
   & \textbf{mean\_clinical \_bert\_sim (M11)} & 0.887 &  0.787 & 0.0 & 228.817 & [ 0.886 , 0.889 ] \\ 
   & \textbf{mean\_legal\_bert\_sim (M8)} & 0.857 & 0.708 & 0.0 & 166.797 & [ 0.855 , 0.859 ] \\ 
   & mean\_scibert\_sim (M10) & 0.714 & 0.609 & 2.468e-24 & 10.347 & [ 0.712 , 0.717 ] \\ 
   & mean\_specter\_sim (M9) & 0.760 & 0.683 & 8.244e-154 & 29.573 & [ 0.756 , 0.764 ]  \\ 
    & mean\_political\_bert\_sim (M7) & 0.780 & 0.591 & 0.0 & 53.434 & [ 0.855 , 0.859 ]  \\ 
    & mean\_sentence\_t5\_xl\_sim (M6) & 0.731 & 0.642 & 4.273e-102 & 23.090 &  [ 0.729 , 0.734 ] \\ 
    & mean\_bert\_large\_uncased\_sim (M2) & 0.775 & 0.612 & 0.0 & 50.625 & [ 0.772 , 0.778 ]   \\
  \midrule
   \multirow{3}{*}{Case 2} & mean\_biobert\_base\_cased\_v12\_sim (M12) & 0.807 & 0.786 &  0.0 & 48.193 & [ 0.805 , 0.808 ]\\ 
   & mean\_clinical\_bert\_sim (M11) &  0.805 & 0.787 & 0.0 & 189.691 & [ 0.804 , 0.806 ]\\ 
   & mean\_legal\_bert\_sim  (M8) &  0.755 & 0.708 & 0.0 & 59.459 & [ 0.753 , 0.756 ]\\ 
 \midrule

  \multirow{3}{*}{Case 3} &  mean\_biobert\_base\_cased\_v12\_sim (M12) &  0.787 & 0.786 & 0.0 & 216.246 & [ 0.786 , 0.788 ]\\ 
   & mean\_clinical\_bert\_sim (M11) &  0.787 & 0.787 & 0.0 & 254.249 & [ 0.787 , 0.788 ]\\ 
   & mean\_legal\_bert\_sim (M8) &  0.709 & 0.708 & 1.141e-51 & 15.569 & [ 0.708 , 0.71 ]\\ 
  \bottomrule
\end{tabular}
\label{tab:inferential statistical analysis}
\end{table*}

From the above statistical analysis and observations, we conclude that pretrained LLMs used in \textbf{\textit{cluster4}} (M11, M12) and \textbf{\textit{cluster3}} (M2, M6, M7, M8, M9, M10) can potentially be used in the future analysis of the research and policy documents. Moreover, domain-specific pertained models (e.g., mean\_clinical\_bert\_sim (M11), and mean\_biobert\_sim (M12)) can provide better outcomes in research involving text analysis in the policy domain (\textbf{RQ1}). Again, higher relevance scores returned by the domain-specific models indicate the higher relevance of research articles cited in policy documents in general. Moreover, the closeness of the means and medians of the relevance scores distribution observed in the box-whisker plots (figure \ref{fig:case1}, figure \ref{fig:case2}, and figure \ref{fig:case3}) and in Table \ref{tab:inferential statistical analysis} provides evidence of the data being nearly normally distributed. This highlights the potential of these models in quantifying the relevance between research and citing policy documents, which could be valuable for all stakeholders and contribute to enhancing trust and facilitating the selection of relevant research articles as research evidence in policymaking (\textbf{RQ2}). Furthermore, we observed nearly identical relevance score distributions for all models in {\verb|Case 2|} and {\verb|Case 3|} across all analyses. This finding suggests that the 'full-text' of research articles does not seem to provide significant additional information beyond what is conveyed by the 'abstract'. Consequently, one implication is that the current approach to seeking research evidence primarily relies on the 'abstract' as it sums up the main theme of any research work for general audience with minimal to no technical jargons or in-depth systematic complexity and analysis. Moreover, the limited time available to policymakers, coupled with the challenges of extracting research evidence from the vast volume of studies by reviewing entire texts, may contribute to such observations. However, it is worth exploring in future research to what extent different sections of a research article contribute to policy drafts, as well as the types of research evidence utilized and their specific purposes. Further experiments and analysis based on this dataset is discussed in two other related works in \cite{b42, b43}.  
\section{Conclusion}
Investigating how research evidence is used in policy documents is integral to bridging the gap between academic knowledge and its practical application in society. It not only facilitates a better understanding of the societal relevance of academic work but also empowers researchers to actively contribute to developing and implementing policies that address pressing societal issues. In this study, we investigated the viability of open-source pretrained large language models (LLMs) in assessing the research relevance of youth policies. We demonstrated the process of collecting and analyzing extensive datasets comprising research and policy documents and illustrated how LLMs, in conjunction with statistical analysis, can aid in discerning the potential impact of research on policy. Looking ahead, we plan to expand this work by exploring tasks such as research impact prediction and experimenting with additional models to derive more generalized conclusions across the domain. Furthermore, we aim to develop a metric that can be utilized for future evaluations of research impact on policy. The data and code of this work is made available in \url{https://github.com/JannatMokarrama07/Research-Policy-Relevance} 

\section*{Acknowledgment}
This work is supported in part by NSF Grant No. 2022443. We thank Terry Bucknell and his team for providing access to \textit{Overton} to conduct this research.


\end{document}